\documentclass{article}
\usepackage{graphicx} 

\usepackage[ruled,vlined]{algorithm2e}
\SetKwBlock{InitBlock}{Initialization ($i=0$)}{end}
\SetKwBlock{UpdateBlock}{Sequential update ($i\mapsto i+1$)}{end}

\usepackage{setspace}
\usepackage{graphicx}
\usepackage{xcolor}
\usepackage{enumitem}
\usepackage[export]{adjustbox}
\usepackage{amsmath,amssymb,amsfonts}%
\usepackage[margin=2.5cm]{geometry}
\usepackage{float}
\usepackage[ruled,vlined]{algorithm2e}
\usepackage{multirow}
\usepackage{array}
\usepackage{hyperref}

\title{Hierarchical Bayesian estimation for continual learning during model-informed precision dosing}

\author{Franziska Thoma$^{1,*}$, Niklas Hartung$^{1,*}$, Manfred Opper$^{2}$, Wilhelm Huisinga$^{1}$}

\date{%
\small
    $*$ joint first authors\\
    $^1$Institute of Mathematics, University of Potsdam, Germany\\%
    $^2$Institute of Software Engineering and Theoretical Computer Science, TU Berlin, Germany\\[2ex]%
    \today
}

\definecolor{up-blue}{RGB}{0,48,94} 
\definecolor{mathnat-blue}{RGB}{0,128,181} 
\definecolor{MyDarkGreen}{RGB}{0,100,0}

\begin{document}
	
	\maketitle

	\begin{abstract}
	Model informed precision dosing (MIPD) is a Bayesian 
	framework to individualize drug therapy based on prior knowledge and patient-specific monitoring data.
	Typically, prior knowledge results from controlled clinical trials with a more homogeneous patient population compared to the real-world patient population underlying the data to be analysed. 
	Thus, devising algorithms that can learn the distribution underlying the real-world patient population from patient-specific monitoring data is of key importance.	
	Formulating continual learning in MIPD as a hierarchical Bayesian estimation problem,
	we here investigate different algorithms for the resulting marginal posterior inference problem in a pharmacokinetic context and for different data sparsity scenarios. 
	As an accurate but computationally expensive reference method, a Metropolis-Hastings algorithm adapted to the hierarchical setting was used.
	Furthermore, several sequential algorithms were investigated: a nested particle filter, a newly developed simplification termed single inner nested particle filter, as well as an approximative parametric method that allows to use Metropolis-within-Gibbs sampling.
	The single inner nested particle filter showed the best compromise between accuracy and computational complexity. 
	Applications to more challenging MIPD scenarios from cytotoxic chemotherapy and anticoagulation initiation therapy are ongoing. 	
	\end{abstract}

	\clearpage
	
	\section{Introduction}
	
	Model-informed precision dosing (MIPD) is a quantitative framework to individualise drug therapy based on prior knowledge and patient-specific monitoring data \cite{Keizer2018,Peck2020,Kluwe2020}.
	Prior information is usually obtained from clinical trials, in the form of nonlinear mixed-effects models estimated from the trial data, and parameter distributions representing inter-individual variability are used as priors in a Bayesian setting.
	In contrast, monitoring data are not obtained in a trial setting but in clinical practice (real-world data).
	When trial and clinical populations differ, the prior parameter distribution may lead to biased predictions during MIPD.
	Through the incorporation of patient-specific data, Bayesian estimation allows to alleviate this bias \cite{Maier2020}, but only when enough monitoring data are available --- initially, dosing recommendations may still be biased.
	For unbiased dose recommendations even at therapy initiation, it is necessary to update the prior parameter estimation between individuals, a problem we will refer to as \emph{continual learning}, as it is used in the machine learning literature \cite{Wang2024}.
	Continual learning can be achieved by a hierarchical Bayesian framework in which the prior parameter distribution on the individual level is not fixed, but itself a random quantity. 
	As a consequence, the fully observable problem becomes a partially observed problem, since only the data (lowest level of the hierarchy), but not the individual parameters (intermediate level of the hierarchy) are observed.
	The resulting statistical problem is thus that of marginal posterior inference.
	The pseudo-marginal Metropolis-Hastings algorithm generalizes classical Metropolis-Hastings to the partially observed (including the hierarchical Bayesian) setting and solves this problem with theoretical guarantees \cite{Beaumont2003,Andrieu2009}.
	Being a batch algorithm, it does not approach the estimation problem sequentially, rendering it costly and potentially impractical in practice.
	We have previously investigated a parametric approximation combined with Metropolis-within-Gibbs sampling in a realistic neutropenia model where pseudo-marginal Metropolis-Hastings is expected to be computationally too expensive \cite{Maier2021}. 
         In a sparse data setting, however, lack of convergence was observed. One aim of the present study is to investigate, whether the observed lack of convergence is attributed to the sparse data setting or due to the simplifying assumptions underlying the approach used, and whether alternative approaches with less restrictive assumptions resolve the problem.
	Particle-based methods have been proposed to the hierarchical setting \cite{Chopin2012,Crisan2018}, but have not yet transferred to the MIPD setting.
	
	In this work, we therefore systematically evaluate different sequential algorithms for the hierarchical Bayesian estimation problem in continued learning and compare them to the pseudo-marginal Metropolis-Hastings algorithm that is considered the reference approach. 
	Besides pseudo-marginal Metropolis-Hastings, the parametric approximation of Metropolis-within-Gibbs and the nested particle filter, we also evaluate a newly developed simplified version of the nested particle filter called single inner nested particle filter.
	Here, we investigate a simplified setting in which all algorithms are feasible, and both runtime and accuracy are evaluated in four scenarios of different data sparsity.

	\section{Statistical framework and algorithms}
	
	\subsection{Nonlinear mixed-effect modelling of clinical trial data}

	Model-informed precision dosing leverages a model estimated from prior clinical trial data, and hence we start by describing this setting.
	
	\paragraph{Data.}
		Measurements (plasma concentrations, biomarkers) are obtained from $N_\text{prior}$ individuals.
		The $j$-th measurement for individual $i$ (at time point $t_{ij}$) is denoted $y_{ij}$.
		The collection of all $n_{i}$ measurements for individual $i$ is denoted by $\mathbf{y}_{i} = (y_{i1},...,y_{in_{i}})$, and all observations for individuals 1 to $N$ by $\mathbf{y}_{1:N} = (\mathbf{y}_{1},...,\mathbf{y}_{N})$.
		One or several doses are given; to simplify notation, we assume here that each individual receives a single (possibly individual-specific) dose $d_{i}$ at time $t=0$. 
		In addition, patient-specific covariates (e.g., body weight, renal function) could be integrated, which are omitted here, also to simplify notation.

	\paragraph{Structural model.}
		A pharmacokinetic (or pharmacokinetic-pharmacodynamic) model describes disposition (and possibly effect) of the drug after dosing, formulated as a parametrized system of ordinary differential equations for the (non-observable) state $x_i(t) = x_{i}(t; \theta_{i}, d_{i})$ of individual $i$,
		\begin{equation}
		\label{eq:StructuralModel}
		\frac{dx_{i}}{dt}(t) = f(x_{i}(t), \theta_{i}), \qquad x_{i}(0) = x_0(\theta_{i}, d_{i}),
		\end{equation}
		with drug dosing $d_{i}$ entering the ODE system via the initial conditions. 
		The patient state $x_{i}$ can be indirectly observed via an observation function $h(x_{i})$. 

	\paragraph{Statistical model.}	
		To represent inter-individual variability, the parameters of the structural model are not fixed, but rather vary within the population according to a \emph{variability model} 
		\begin{equation}
		\label{eq:VariabilityModel}
		\theta_{i}\sim_\text{iid} p(\cdot;\zeta)
		\end{equation}
		with hyperparameters $\zeta$ to be estimated.
		Note that, in MIPD, inter-individual variability will be interpreted as uncertainty for a new individual.
		The solution $x_i(t) = x_{i}(t; \theta_{i}, d_{i})$ to Eq.~\eqref{eq:StructuralModel} is then related to the data $\mathbf{y}_{i}$ via an \emph{observation model} 
		\begin{equation}
		\label{eq:ObservationModel}
		[\mathbf{y}_{i}|\theta_{i}] \sim p_{i}(\cdot|\theta_{i}) := p(\cdot;h_{i}(x_{i},\theta_{i})).
		\end{equation}
		such as, for example, $[y_{ij}|\theta_{i}] \sim_\text{iid} \mathcal{N}\Big( h(x_{i}(t_{ij}; \theta_{i}, d_{i})), \sigma^{2}\Big)$, $j=1,...,n_{i}$.
		Furthermore, the data $\mathbf{y}_{1},...,\mathbf{y}_{N_\text{trial}}$ for different individuals are assumed to be conditionally independent given their respective parameters $\theta_{1},...,\theta_{N_\text{trial}}$.

	The model comprised of Eqs.~\eqref{eq:VariabilityModel}-\eqref{eq:ObservationModel} is a \emph{nonlinear mixed-effect (NLME) model}.
	Importantly, the individual parameters $\theta_{1},...,\theta_{N_\text{trial}}$ are latent -- instead of estimating the individual parameters themselves, a parameter (point) estimate $\hat\zeta_\text{NLME}$ for their distribution is derived.
	The computational challenges in NLME modelling have been extensively discussed and several software packages for  parameter estimation are available \cite{Sheiner1980,Delyon1999,Comets2017}.
	
	\subsection{Hierarchical Bayesian framework for model-informed precision dosing}

	We now come to the MIPD setting, where it is assumed that an estimate $\hat\zeta_\text{NLME}$ in an NLME model  has already been determined (see previous section).
	The aim of MIPD is to inform a dosing decision for a new individual based on the NLME model and parameter estimates.
	
	In the non-hierarchical setting, the variability model \eqref{eq:VariabilityModel} is used as a prior $p(\theta;\hat\zeta_\text{NLME})$ for a new individual's parameters $\theta$ in a Bayesian framework.
	This approach allows to inform the dosing of a new individual from the clinical trial data.
	However, it does \emph{not} allow to learn from already observed \emph{monitoring} data, i.e., to update the prior based on the already treated individuals in MIPD. The ability to update the prior distribution over time is particularly relevant in the expected scenarios, in which the clinical trial data used to infer $\zeta_\text{NLME}$ markedly differ from the therapeutic drug monitoring data underlying MIPD in real-world settings. 
	To incorporate the possibility to update the prior given monitoring data, the Bayesian framework can be extended by addition layer of hierarchy. 
	Instead of assuming a fixed value $\zeta = \hat\zeta_\text{NLME}$, a prior distribution $p$ is given for $\zeta$. 
	The resulting hierarchical Bayesian framework is specified on three levels: 
	
	\paragraph{Population level.}
	A prior $p$ is chosen for the population parameters $\zeta$, centred in the nonlinear mixed-effects model estimates $\hat\zeta_\text{NLME}$. Its width is typically related to the uncertainty in these estimates---either reflecting statistical uncertainty of $\hat\zeta_\text{NLME}$, or including an inflation factor anticipating a change in distribution to be expected in the real-world population.
	
	\paragraph{Individual level.}
	Subsequently, monitoring data are collected from patients, usually in a sequential manner (one patient at a time).
	The individual parameters $\theta_{i}$, $i=1,...,N$ with $N$ denoting the number of patients, are assumed conditionally i.i.d. given the population parameters $\zeta$, with density $p(\theta_{i}|\zeta)$ from \eqref{eq:VariabilityModel}.
	
	\paragraph{Observation level.} 
	The data $\mathbf{y}_{i}$ for individual $i$ are assumed to follow the observation model $p_{i}(\mathbf{y}_{i}|\theta_{i})$ from \eqref{eq:ObservationModel}.
	The conditional independence assumption in \eqref{eq:ObservationModel} is extended to also include the population parameters $\zeta$, i.e., it is assumed that population parameters $\zeta$ and individual data $\mathbf{y}_{1},...,\mathbf{y}_{N}$ are  conditionally independent given the individual parameters $\theta_{1},...,\theta_{N}$ , which implies in particular that 
		\begin{equation}
		\label{eq:conditional-independence}
		p(\mathbf{y}_{1:N}|\zeta,\theta_{1:N}) = \prod_{i=1}^{N} p_{i}(\mathbf{y}_{i}|\theta_{i}).
		\end{equation}	
	
	\paragraph{Marginal posterior.}
	In this setting, the information on $\zeta$ contained in monitoring data $\mathbf{y}_{1:N}$ is represented via the (marginal) posterior distribution of the population parameters $\zeta$ given all observations,
	\begin{equation}
	\label{eq:marginal-posterior}
	p(\zeta|\mathbf{y}_{1:N}) \propto p(\mathbf{y}_{1:N}|\zeta)\cdot p(\zeta).
	\end{equation}
	Using conditional independence \eqref{eq:conditional-independence}, the (marginal) likelihood simplifies to 
	\begin{equation}
	\label{eq:marginal-likelihood}
	p(\mathbf{y}_{1:N}|\zeta) = \prod_{i=1}^{N} \int p_{i}(\mathbf{y}_{i}|\theta)p(\theta|\zeta) d\theta.
	\end{equation}
	Note that the calculation of $p_{i}(\mathbf{y}_{i}|\theta)$ involves solving a system of ODEs, and that the marginal likelihood usually cannot be computed in closed form due to the integral in $\theta$.

	\subsection{Algorithms for marginal posterior inference}
	
	In this section, we  outline several algorithms for hierarchical Bayesian inference. 
	All of these methods compute a sample from the marginal posterior (full Bayesian inference), 
	unlike point-based methods such as maximum-a-posteriori estimation which will not be considered.
	
	\subsubsection*{Pseudo-marginal Metropolis-Hastings with importance sampling}
	
	The pseudo-marginal Metropolis-Hastings algorithm is an extension of the classical Metropolis-Hastings algorithm \cite{Metropolis1953,Hastings1970} suitable for the hierarchical setting.
	At step $l$, a new sample $\zeta^{\ast}$ is proposed according to a proposal distribution $q$ that depends on the parameter $\zeta^{(l-1)}$ of the previous step, i.e. $\zeta^{\ast} \sim q(\cdot|\zeta^{(l-1)})$.
	In classical Metropolis-Hastings, this proposal is then accepted with probability 	
	\begin{equation}
	\label{eq:acceptance-standard}
	\alpha = \min\left(1,\; \frac{p(\zeta^{\ast}|\mathbf{y}_{1:N})}{p(\zeta^{(l-1)}|\mathbf{y}_{1:N})}\frac{q(\zeta^{(l-1)}|\zeta^{\ast})}{q(\zeta^\ast|\zeta^{(l-1)})}\right)
	\end{equation}
	However, this calculation involves a potentially high-dimensional and hence computationally expensive integration in the marginal likelihood \eqref{eq:marginal-likelihood}.
	Pseudo-marginal Metropolis-Hastings replaces the exact marginal likelihood terms in \eqref{eq:acceptance-standard} by an unbiased estimator, here in form of a Monte Carlo approximation 
	\begin{equation}
	\label{eq:pseudo-marginal-likelihood}
	\int p_i(\mathbf{y}_{i}|\theta)p(\theta|\zeta) d\theta \approx \frac{1}{M}\sum_{m=1}^{M} p_i(\mathbf{y}_{i}|\theta^{(m)}), \qquad \text{with } \theta^{(m)}\sim_\text{iid}p(\cdot|\zeta).
	\end{equation}
	It has been shown that the stationary distribution of the Markov chain is unaffected by this approximation \cite{Andrieu2009,Beaumont2003}, and hence pseudo-marginal Metropolis-Hastings can be considered as a reference algorithm for benchmarking other hierarchical Bayesian methods.
	We use a slight modification of pseudo-marginal Metropolis-Hastings that includes importance sampling to ensure that the same vector $\theta^{(1)},...,\theta^{(M)}$ is used in both the numerator and denominator of the acceptance ratio \eqref{eq:acceptance-standard}.
	In contrast to plain pseudo-marginal Metropolis-Hastings, pseudo-marginal Metropolis-Hastings with importance sampling does not have the theoretical guarantee of unbiasedness, but it showed superior performance in preliminary numerical experiments (faster convergence).
	The detailed steps are given in Algorithm~\ref{algo:pm-MH-is}.	
	
	\bigskip
	
	\begin{algorithm}[H]
\label{algo:pm-MH-is}
\caption{Pseudo-marginal Metropolis-Hastings with importance sampling (batch)}
\SetAlgoLined
\KwIn{Data for all individuals $\mathbf{y}_{1:N}$, initial guess $\zeta^{(0)}$, length of Markov chain $L$, Monte Carlo sample size $M$, prior distribution $p$ for $\zeta$, proposal distribution $q(\cdot|\cdot)$ for $\zeta$}
\KwOut{Samples $\zeta^{(1)},...,\zeta^{(L)}$ from the marginal posterior $p(\cdot | \mathbf{y}_{1:N})$}
\BlankLine

\For{$l = 1$ \KwTo $L$}{
    Propose $\zeta^{\ast} \sim q(\cdot | \zeta_{l-1})$\;
    Sample $\theta^{(1:M)} \sim_\text{iid} p(\cdot|\zeta^{\ast})$\;
    
    \For{$i=1$ \KwTo $N$}{
    	\For{$m=1$ \KwTo $M$}{
    		Evaluate structural model to obtain $p_i(\mathbf{y}_{i}| \theta^{(m)})$\;
    	}
	Approximate marginal likelihood (for individual $i$):
	\begin{align*}
	\hat p_i(\mathbf{y}_{i}|\zeta^{\ast}) &= \frac{1}{M} \sum_{m=1}^{M}p_i(\mathbf{y}_{i}| \theta^{(m)})\\
	\qquad\qquad\hat p_i(\mathbf{y}_{i}|\zeta_{l-1}) &= \frac{1}{M} \sum_{m=1}^{M}p_i(\mathbf{y}_{i}| \theta^{(m)})\frac{p(\theta^{(m)}|\zeta_{l-1})}{p(\theta^{(m)}|\zeta^{\ast})} \qquad\qquad\qquad \texttt{// Importance sampling}
	\end{align*}
    }
    Assemble marginal likelihood (for all individuals):
    \[
    \hat p(\mathbf{y}_{1:N}|\zeta^{\ast}) = \prod_{i=1}^{N} \hat p_i(\mathbf{y}_{i}|\zeta^{\ast}),
    \qquad
    \hat p(\mathbf{y}_{1:N}|\zeta^{(l-1)}) = \prod_{i=1}^{N} \hat p_i(\mathbf{y}_{i}|\zeta_{l-1});
    \]
    With probability
    \[
    \alpha = \min\left(1, \, \frac{\hat p(\mathbf{y}|\zeta^{\ast}) p(\zeta^{\ast}) q(\zeta^{(l-1)} | \zeta^{\ast})}{\hat p(\mathbf{y}|\zeta^{(l-1)})p(\zeta_{l-1}) q(\zeta^{\ast} | \zeta^{(l-1)})}\right),
    \]
    set $\zeta_{l} = \zeta^{\ast}$, otherwise $\zeta_{l} = \zeta^{(l-1)}$\;
}
\end{algorithm}

	\bigskip

	\subsubsection*{Nested and single inner nested particle filters}
	
	Particle filters are a class of sequential methods for full Bayesian inference in general setting, i.e., not relying on any Gaussian distributional assumptions \cite{Arulampalam2002}.
	Commonly used in weather prediction, they have also been introduced into the MIPD field recently \cite{Elfring2021,Maier2021}.	

	We first describe the classical particle filter in a non-hierarchical Bayesian setting, 
	i.e. Eqs.~\eqref{eq:VariabilityModel}-\eqref{eq:ObservationModel} with $N=1$ and for a fixed $\zeta\ (= \hat\zeta_\text{NLME})$. Since not relevant in this case, we simplify notation and drop both $i$ and $\zeta$ from the notation here. 	
	As sequential methods, particle filters rely on an iterative reformulation of Bayes' formula, 
	\begin{equation}
	\label{eq:individual-posterior-sequential}
	p(\theta| y_{1:(j+1)}) \propto p(y_{j+1}| \theta)\cdot p(\theta| y_{1:j}).
	\end{equation}
	At each step in the iteration, an approximation of the posterior $p(\theta| y_{1:j})$ is given by the empirical distribution in an ensemble of particles $\theta^{(1:S)}_{j}=(\theta^{(1)}_{j},...,\theta^{(S)}_{j})$ with corresponding weights $w^{(1:S)}_{j}=(w^{(1)}_{j},...,w^{(S)}_{j})$.
	The iterative formulation of Bayes' formula	is then used to update the particle weights according to the likelihood,
	\begin{equation}
	\label{eq:inner-update}
	w^{(s)}_{j+1} = w^{(s)}_{j} \cdot p(y_{j+1}|\theta^{(s)}_{j}),\qquad s\in\{1,...,S\}.
	\end{equation}
	If particle weights $w^{(1:S)}_{j+1}$ concentrate too much on a low number of particles, measured via a decrease in the effective sample size 
	\[
	S_\text{eff}\big(w^{(1:S)}_{j+1}\big) := \sum_{s=1}^{S} \big(w^{(s)}_{j+1}\big)^{2},
	\]
	a resampling/rejuvenation step can be performed after any iteration: 
	\begin{itemize}
	\item \emph{resampling:} sample $S$ new i.i.d.~particles $\theta^{(1:S)}_{j+1}$ from $\sum_{s=1}^{S}w^{(s)}_{j+1}\delta_{\theta^{(s)}_{j}}$ and reset $w^{(s)}_{j+1}=\frac{1}{S}$; 
	\item \emph{rejuventation:} perturb each newly sampled particle $\theta_{j+1}^{(s)}$ by random additive noise $\varepsilon^{(s)}\sim_\text{iid}\mathcal{N}(0,\sigma_\text{rej}^{2})$, with rejuvenation variance $\sigma_\text{rej}^{2}$ being a hyperparameter of the algorithm.
	\end{itemize}
	Otherwise, the particles stay the same, i.e., $\theta^{(s)}_{j+1}=\theta^{(s)}_{j}$ for all $s$.
	
	In the hierarchical setting, the iterative formulation of Bayes' formula yields 
	\begin{equation}
	\label{eq:marginal-posterior-sequential}
	p(\zeta|\mathbf{y}_{1:(i+1)}) \propto p(\mathbf{y}_{i+1}|\zeta)\cdot p(\zeta|\mathbf{y}_{1:i});
	\end{equation}
	note the difference to \eqref{eq:individual-posterior-sequential}.
	To deal with this hierarchical Bayesian setting, the \emph{nested particle filter} has been proposed \cite{Chopin2012,Crisan2018}.
	In this algorithm, different particle ensembles (outer and inner) are used to represent the population ($\zeta$) and individual ($\theta$) levels: 
	the outer weighted particle ensemble $(v_{i}^{(1)},\zeta^{(1)}_{i}),...,(v_{i}^{(R)},\zeta^{(R)}_{i})$ represents $p(\zeta|\mathbf{y}_{1:i})$, and for each $\zeta^{(r)}_i$, a corresponding inner weighted particle ensemble $(w_{r}^{(1)},\theta_{r}^{(1)}),...,(w_{r}^{(S)},\theta_{r}^{(S)})$ represents $p(\theta_{i}|\mathbf{y}_{i+1},\zeta^{(r)}_{i})$.
	Note that, as in the non-hierarchical setting, the inner ensemble can also be updated sequentially at each individual datapoint $y_{i+1,j}$; 
	here, we omit this level of detail (i.e., weights $w_{r}^{(s)}$ correspond to the final result after accounting for all individual data $\mathbf{y}_{i+1} = (y_{i+1,1},...,y_{i+1,n_{i}})$).
	After calculating the inner weights, the outer weights are updated, 
	\begin{equation}
	\label{eq:outer-update-individually}
	v_{i+1}^{(r)} \propto v_{i}^{(r)} \cdot \sum_{s=1}^{S} w_{r}^{(s)}.
	\end{equation}
	Resampling/rejuvenation can be applied for the outer particle ensemble in the same way described above for the non-hierarchical particle filter.
	In contrast, for the inner particle ensemble, resampling cannot be used in the same way, since only the inner particle \emph{weights} (not the particles themselves) inform the outer level. 
	The detailed steps for the nested particle filter are described in Algorithm~\ref{algo:nPF}.
	
	\bigskip

	\begin{algorithm}[H]
\label{algo:nPF}
\caption{Nested particle filter (sequential)}
\SetAlgoLined

\KwIn{Prior distribution $p$ for $\zeta$, data for all individuals $\mathbf{y}_{1:N}$, outer particle ensemble size $R$, inner particle ensemble size $S$}
\KwOut{Weighted outer particle ensemble $(v^{(1)}_{N},\zeta^{(1)}_{N}),...,(v^{(R)}_{N},\zeta^{(R)}_{N})$ such that $\sum_{r=1}^{R} v^{(r)}_{N}\delta_{\zeta^{(r)}_{N}} \approx p(\cdot | \mathbf{y}_{1:N})$}
\BlankLine
\InitBlock{
Sample initial outer particle ensemble $\zeta_{0}^{(1:R)}$ from the prior $p$\;
Set $v^{(r)}_{0} = \frac{1}{R}$ for all $r$\;
}
\BlankLine
\UpdateBlock{
\BlankLine
\KwIn{Weighted outer particle ensemble $(v^{(1)}_{i},\zeta^{(1)}_i),...,(v^{(R)}_{i},\zeta^{(R)}_{i})$ such that $\sum_{r=1}^{R} v^{(r)}_{i}\delta_{\zeta^{(r)}_{i}} \approx p(\cdot | \mathbf{y}_{1:i})$, data for a new individual $\mathbf{y}_{i+1}$}
\KwOut{Weighted outer particle ensemble $(v^{(1)}_{i+1},\zeta^{(1)}_{i+1}),...,(v^{(R)}_{i+1},\zeta^{(R)}_{i+1})$ such that $\sum_{r=1}^{R} v^{(r)}_{i+1}\delta_{\zeta^{(r)}_{i+1}} \approx p(\cdot | \mathbf{y}_{1:(i+1)})$}
\BlankLine
\For{$r=1$ \KwTo $R$}{
	Sample inner particle ensemble $\theta_{r}^{(1:S)}\sim_\text{iid} p(\cdot| \zeta_{i}^{(r)})$\;
	\For{$s=1$ \KwTo $S$}{
		Calculate inner weights $w_{r}^{(s)} = p(\mathbf{y}_{i+1}| \theta_{r}^{(s)})$\;
	} 
	Update outer weights $v^{(r)}_{i+1} \propto v^{(r)}_{i} \cdot \sum_{s=1}^{S}w_{r}^{(s)}$\;
	}
\BlankLine
\uIf{$S_\text{eff}\,\big(v^{(1:R)}_{i+1}\big) \ge \text{threshold}$}{
$\zeta^{(r)}_{i+1} = \zeta^{(r)}_{i}$ for all $r$\;
}\Else(\tcp*[f]{(Outer) resampling \& rejuvenation}){
$\zeta^{(1)}_{i+1},\ldots,\zeta^{(R)}_{i+1} \sim_\text{iid} \sum_{r=1}^{R} v^{(r)}_{i+1}\delta_{\zeta^{(r)}_{i}}$ plus rejuvenation\\
$v^{(r)}_{i+1} = \frac{1}{R}$ for all $r$\; 
}
}
\end{algorithm}

	\bigskip
	
	The nested particle filter described above requires to sample $R\cdot S$ inner particles in each iteration, but these may be redundant and thereby lead to excess computational effort. 
	This is in particular to be expected if the observations and the hyperparameter $\zeta$ are independent given the individual parameters, Eq.~\eqref{eq:conditional-independence}.
	Based on this observation, we introduce here a variant termed \emph{single inner nested particle filter}.
	This algorithm avoids the computational burden of the standard nested particle filter by only simulating a single inner ensemble $\theta_\text{ref}^{(1)},...,\theta_\text{ref}^{(S)}$ sampled from a suitably chosen reference outer particle $\zeta_\text{ref} = \texttt{summary}(\zeta_{i}^{(1)},...,\zeta_{i}^{(R)})$, chosen to cover the entire range of particles produced by the nested particle filter.
	The single inner ensemble is then used to simultaneously update the weights of all $R$ outer particles,
	accounting for the choice of reference particle via an importance ratio:
	\begin{equation}
	\label{eq:outer-update-importance-ratio}
	v_{i+1}^{(r)} \propto v_{i}^{(r)}  \sum_{s=1}^{S} \left(w_\text{ref}^{(s)}\cdot \frac{p(\theta_\text{ref}^{(s)}| \zeta^{(r)})}{p(\theta_\text{ref}^{(s)} | \zeta_\text{ref})}\right).
	\end{equation}
	
\begin{figure}
		\centering
		\includegraphics[width=.9\textwidth,trim=0 0 0 0,clip]{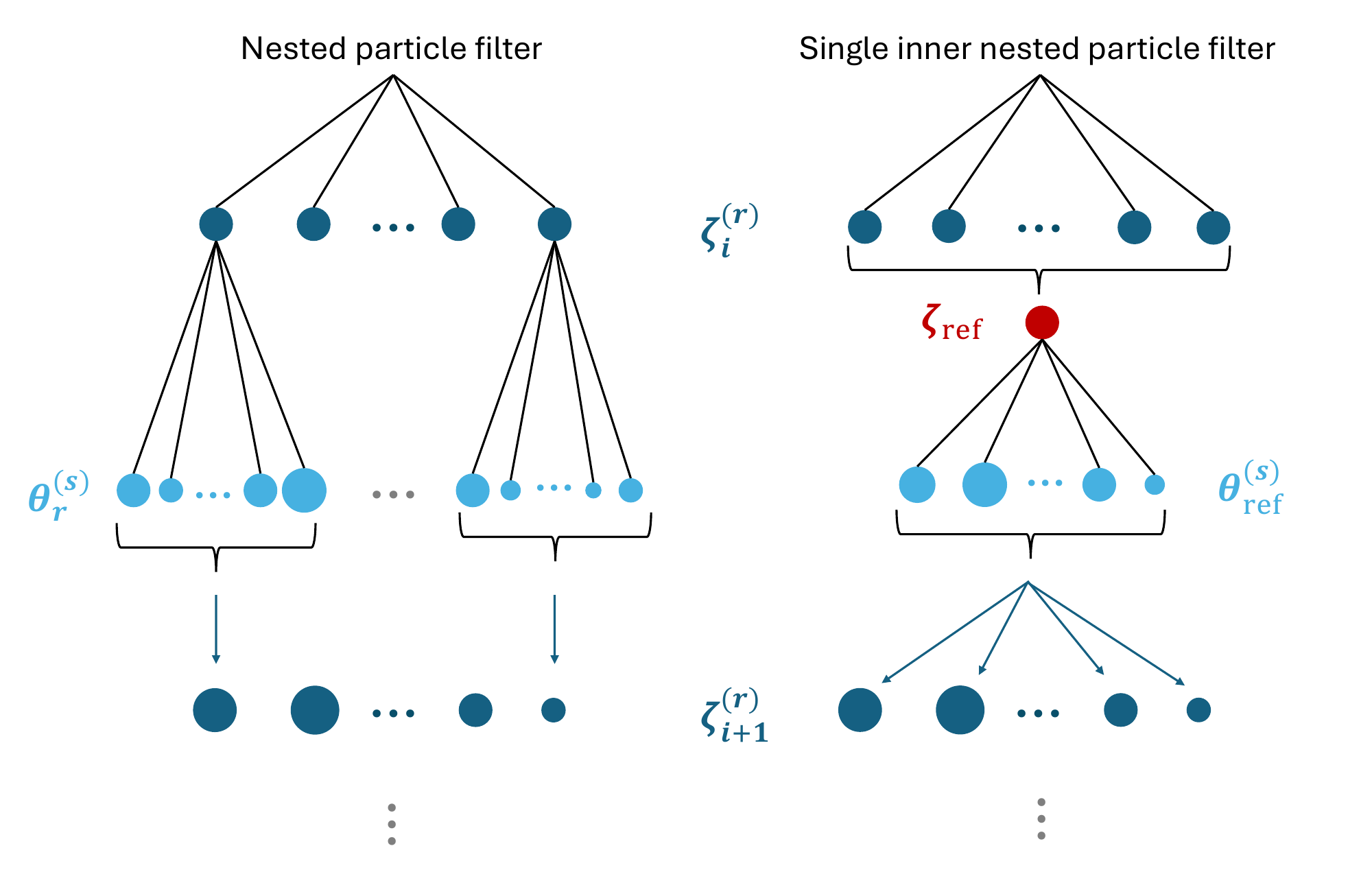}
		
		\caption{\label{fig:nPFsinPF}
		Visual explanation of the two variants of the nested particle filter for hierarchical Bayesian inference.
		Nested particle filter (left): For each particle $\zeta^{(r)}_{i}$ in the current outer ensemble (dark blue, top), a corresponding inner particle ensemble $\theta_{r}^{(1:S)}$ is initialized and weights $w_{r}^{(1:S)}$ are derived based on the observation model (light blue, area corresponds to weight), cf.~\eqref{eq:inner-update}. 
		Subsequently, for each outer particle $\zeta^{(r)}_{i}$, its weight $v_{i}^{(r)}$ is updated based on $w_{r}^{(1:S)}$  (dark blue, bottom), cf.~\eqref{eq:outer-update-individually}. 
		In contrast, the single inner nested particle filter (right) generates a particle representation $\zeta_\text{ref}$ (red), which initializes only one reference inner particle ensemble $\theta_\text{ref}^{(1:S)}$. 
		All outer weights $v_{i}^{(1:R)}$ are then updated simultaneously using the reference inner weights $w_\text{ref}^{(1:S)}$ adjusted to each outer particle by the individual importance ratio \eqref{eq:outer-update-importance-ratio}.
		}
	\end{figure}

	The (conventional) nested and the single inner nested PF algorithms are summarized graphically in Fig.~\ref{fig:nPFsinPF}; the detailed steps for the single inner nested PF are given in Algorithm~\ref{algo:sinPF}.
	
\bigskip

\begin{algorithm}[H]
\label{algo:sinPF}
\caption{Single inner nested particle filter (sequential)	
}
\SetAlgoLined

\KwIn{Prior distribution $p$ for $\zeta$, data for all individuals $\mathbf{y}_{1:N}$, outer particle ensemble size $R$, inner particle ensemble size $S$, outer particle summary function}
\KwOut{Weighted outer particle ensemble $(v^{(1)}_{N},\zeta^{(1)}_N),...,(v^{(R)}_{N},\zeta^{(R)}_{N})$ such that $\sum_{r=1}^{R} v^{(r)}_{N}\delta_{\zeta^{(r)}_{N}} \approx p(\cdot | \mathbf{y}_{1:N})$}
\BlankLine
\InitBlock{
Sample initial outer particle ensemble $\zeta^{(1:R)}_{0}$ from the prior $p$\;
Set $v^{(r)}_{0} = \frac{1}{R}$ for all $r$\;
}
\BlankLine
\UpdateBlock{
\BlankLine
\KwIn{Weighted outer particle ensemble $(v^{(1)}_{i},\zeta^{(1)}_i),...,(v^{(R)}_{i},\zeta^{(R)}_{i})$ such that $\sum_{r=1}^{R} v^{(r)}_{i}\delta_{\zeta^{(r)}_{i}} \approx p(\cdot | \mathbf{y}_{1:i})$, data for a new individual $\mathbf{y}_{i+1}$}
\KwOut{Weighted outer particle ensemble $(v^{(1)}_{i+1},\zeta^{(1)}_{i+1}),...,(v^{(R)}_{i+1},\zeta^{(R)}_{i+1})$ such that $\sum_{r=1}^{R} v^{(r)}_{i+1}\delta_{\zeta^{(r)}_{i+1}} \approx p(\cdot | \mathbf{y}_{1:(i+1)})$}
\BlankLine
Generate representation of outer ensemble: $\zeta_\text{ref} = \text{summary}\big(\zeta_{i}^{(1:R)},v_{i}^{(1:R)}\big)$\;
Sample a corresponding inner ensemble $\theta_\text{ref}^{(1:S)}\sim_\text{iid} p(\cdot| \zeta_\text{ref})$\;
\For{$s=1$ \KwTo $S$}{
	Calculate inner weights $w_\text{ref}^{(s)} = p(\mathbf{y}_{i+1}| \theta_\text{ref}^{(s)})$\;
}
\For{$r=1$ \KwTo $R$}{
	Update outer weights $v^{(r)}_{i+1} \propto v^{(r)}_{i} \cdot \sum_{s=1}^{S}\left(w_\text{ref}^{(s)} \cdot \frac{p(\theta_\text{ref}^{(s)} | \zeta^{(r)})}{p(\theta_\text{ref}^{(s)} | \zeta_\text{ref})} \right)$
}
\BlankLine
\uIf{$S_\text{eff}\,(v^{(1:R)}_{i+1}) \ge \text{threshold}$}{
$\zeta^{(r)}_{i+1} = \zeta^{(r)}_{i}$ for all $r$\;
}\Else(\tcp*[f]{(Outer) resampling \& rejuvenation}){
$\zeta^{(1)}_{i+1},\ldots,\zeta^{(R)}_{i+1} \sim_\text{iid} \sum_{r=1}^{R} v^{(r)}_{i+1}\delta_{\zeta^{(r)}_{i}}$\;
$v^{(r)}_{i+1} = \frac{1}{R}$ for all $r$\; 
}
}
\end{algorithm}

\bigskip

	\subsubsection*{Parametric approximation + Metropolis-within-Gibbs sampling}

	All approaches described so far represent the marginal posterior $p(\zeta|\mathbf{y}_{1:N})$ nonparametrically (via a sample/particle ensemble).
	As an alternative, it can be approximated by a parametric class $p(\cdot\;;H)$, $H\in \mathbb{H}$, which we assume to be able to sample from.
	A special sequential algorithm can be designed if this class is conjugate to the distribution of the individual parameters $\theta$, i.e., 
	for $\zeta\sim p(\cdot\;;H)$ there exists $H'=H'(\theta)\in \mathbb{H}$ such that $[\zeta|\theta] \sim p(\cdot\;;H')$.
	In this case, a \emph{Metropolis-within-Gibbs} algorithm can be used in every iteration to sample from $p(\zeta|\mathbf{y}_{1:i})$, coupled with a \emph{parametric approximation} of $p(\zeta|\mathbf{y}_{1:i})$ based on the generated sample.
	
	Gibbs sampling alternates between sampling from conditional distributions to sample from a joint distribution \cite{Geman1984,Gelfand1990}, which is beneficial if sampling from conditional distributions is a structurally simpler problem.
	In Metropolis-within-Gibbs \cite{Gilks1995}, direct sampling from some of these conditional distributions is replaced by Metropolis-Hastings sampling.
	In order to apply Gibbs sampling to the the marginal posterior $p(\zeta|\mathbf{y}_{1:(i+1)})$, the problem is augmented into a joint sampling problem from $p(\zeta,\theta_{i+1}|\mathbf{y}_{1:(i+1)})$; the $\theta_{i+1}$ component from the generated sample can then be simply ignored.
	The conditional distributions to sample from are 
	$p(\zeta^{(l)}|\theta^{(l-1)}_{i+1},\mathbf{y}_{1:(i+1)})$ and $p(\theta^{(l)}_{i+1}|\zeta^{(l)},\mathbf{y}_{1:(i+1)})$, coupled by the iterative update of each other.
	
	As a preparation to Gibbs sampling, and similarly to the single inner nested particle filter, a single ensemble $\theta_\text{ref}^{(1)},...,\theta_\text{ref}^{(S)}$ is first sampled from $p(\cdot|\zeta_\text{ref})$ with a suitably chosen reference value $\zeta_\text{ref}$, derived as a summary of $p(\cdot;H_{i})$, e.g. $\zeta_\text{ref} = \mathbb{E}_{H_{i}}[\zeta]$.
	Subsequently, a weighted ensemble $(w_\text{ref}^{(s)},\theta_\text{ref}^{(s)})$, $s\in\{1,...,S\}$, is determined (non-hierarchical particle filter); $\sum_{s=1}^{S} w_\text{ref}^{(s)}\delta_{\theta_\text{ref}^{(s)}}$ is then an approximation of $p(\,\cdot\,|\zeta_\text{ref},\mathbf{y}_{i+1})$.
	
	Gibbs sampling is then performed as follows:
	\begin{itemize}
	\item Direct sampling $\zeta^{(l)}\sim p(\cdot\,|\theta_{i+1}^{(l-1)},\mathbf{y}_{1:(i+1)})$:
	\begin{itemize}
	\item By the parametric approximation from the previous step, $p(\zeta|\mathbf{y}_{1:i}) = p(\zeta;H_{i})$ for some $H_{i} \in \mathbb{H}$;
	\item Using the sequential Bayes' formula \eqref{eq:marginal-posterior-sequential}, but conditioned on the individual parameters $\theta_{i+1}$,
	\[
	p(\zeta|\theta_{i+1},\mathbf{y}_{1:(i+1)})\propto p(\mathbf{y}_{i+1}|\theta_{i+1},\zeta)\cdot p(\zeta|\theta_{i+1},\mathbf{y}_{1:i});
	\]
	\item By conditional independence, $p(\mathbf{y}_{i+1}|\theta_{i+1},\zeta) = p(\mathbf{y}_{i+1}|\theta_{i+1})$ and hence, this term cancels due to normalization, which yields
	\[
	p(\zeta|\theta_{i+1},\mathbf{y}_{1:(i+1)})\propto p(\zeta|\theta_{i+1},\mathbf{y}_{1:i});
	\]
	\item Due to the conjugacy assumption, $p(\zeta|\theta_{i+1},\mathbf{y}_{1:i}) = p(\zeta;H')$ for some $H'=H'(\theta_{i+1})$, which can be sampled from.
	\end{itemize}
	
	\item 
	Metropolis-Hastings sampling $\theta_{i+1}^{(l)}\sim p(\cdot\,|\zeta^{(l)},\mathbf{y}_{1:(i+1)})$:
	\begin{itemize}
		\item 
		By conditional independence, $p(\theta_{i+1}|\zeta,\mathbf{y}_{1:(i+1)}) = p(\theta_{i+1}|\zeta,\mathbf{y}_{i+1})$ (also note that it is not a \emph{marginal} posterior);
		
		\item A proposal $\theta_{i+1}^{*}\sim \sum_{s=1}^{S} w_\text{ref}^{(s)} \delta_{\theta_\text{ref}^{(s)}}$ is drawn and rejuvenated, constituting an approximate sample from $p(\cdot|\zeta_\text{ref},\mathbf{y}_{i+1})$ (independence sampling);
		\item The  acceptance probability for $\theta_{i+1}^{*}$ (see also Eq.~\eqref{eq:acceptance-standard}) is then given by
		\begin{align*}
		 &\frac{p(\theta_{i+1}^{\ast}|\zeta^{(l)},\mathbf{y}_{i+1})}{p(\theta_{i+1}^{(l-1)}|\zeta^{(l)},\mathbf{y}_{i+1})}\cdot \frac{p(\theta_{i+1}^{(l-1)} | \zeta_\text{ref},\mathbf{y}_{i+1})}{p(\theta_{i+1}^{\ast} | \zeta_\text{ref},\mathbf{y}_{i+1})}\\
		 &= \frac{p(\mathbf{y}_{i+1}|\theta_{i+1}^{\ast})p(\theta_{i+1}^{\ast}|\zeta^{(l)})}{p(\mathbf{y}_{i+1}|\theta_{i+1}^{(l-1)})p(\theta_{i+1}^{(l-1)}|\zeta^{(l)})} \cdot 
		 \frac{p(\mathbf{y}_{i+1}|\theta_{i+1}^{(l-1)})p(\theta_{i+1}^{(l-1)}|\zeta_\text{ref})}{p(\mathbf{y}_{i+1}|\theta_{i+1}^{\ast})p(\theta_{i+1}^{\ast}|\zeta_\text{ref})}\\
		 &=\frac{p(\theta_{i+1}^{\ast}|\zeta^{(l)})}{p(\theta_{i+1}^{(l-1)}|\zeta^{(l)})}\cdot \frac{p(\theta_{i+1}^{(l-1)} | \zeta_\text{ref})}{p(\theta_{i+1}^{\ast} | \zeta_\text{ref})}
		\end{align*}
		The advantage of this formulation is that the observation model \eqref{eq:ObservationModel} need not be evaluated during Gibbs sampling (the resulting expression is independent of $\mathbf{y}_{i+1}$).
	\end{itemize}
	\end{itemize}
	The detailed steps for this algorithm are given in Algorithm~\ref{algo:MwG}.	
	\bigskip

	\begin{algorithm}[H]
\label{algo:MwG}
\SetKwInOut{Requirement}{Requirement}
\caption{Parametric approximation + Metropolis-within-Gibbs sampling (sequential)}
\SetAlgoLined
\KwIn{Data for all individuals $\mathbf{y}_{1:N}$, parametrized prior distribution $p(\zeta)=p(\zeta;H_{0})$, inner ensemble size $S$, length of Markov chain $L$, summary function for $p(\cdot;H)$, estimator for $H$ from observed samples $\zeta^{(1:L)}$}
\KwOut{Hyperparameter $H_{N}$ of the parametric approximation $p(\zeta;H_{N})$ to the posterior $p(\zeta|\mathbf{y}_{1:N})$. 
}
\Requirement{Conjugacy of $p(\cdot;H)$ to $p(\theta|\zeta)$, i.e., for each $H\in\mathbb{H}$ there is a $H'\in\mathbb{H}$ such that $p(\zeta;H') \propto p(\theta|\zeta) p(\zeta;H)$}

\BlankLine
\UpdateBlock{
	\BlankLine
	\KwIn{Hyperparameter $H_{i}$, data for a new individual $\mathbf{y}_{i+1}$}
	\KwOut{Updated hyperparameter $H_{i+1}$}
	\BlankLine
	Generate representation of parametric distribution: $\zeta_\text{ref} = \text{summary}\big(p(\,\cdot\,;H_{i})\big)$\;
	Sample a corresponding inner ensemble $\theta_\text{ref}^{(1:S)}\sim_\text{iid} p(\cdot\,| \zeta_\text{ref})$\;
	\For{$s=1$ \KwTo $S$}{
		Calculate inner weights $w_\text{ref}^{(s)} = p(\mathbf{y}_{i+1}| \theta_\text{ref}^{(s)})$
		\tcp*{Possibly resampling \& rejuvenation}
	}\medskip
	\emph{Gibbs sampling:}\\[3mm]
\For{$l = 1$ \KwTo $L$}{
	\emph{Direct sampling:}
	$\zeta^{(l)} \sim p(\cdot\,|\theta_{i+1}^{(l-1)})$ \tcp*{Requirement is used here}
	\BlankLine
	\emph{Metropolis-Hastings sampling:}
	$\theta_{i+1}^{\ast} \sim \sum_{s=1}^{S} w_\text{ref}^{(s)} \delta_{\theta_\text{ref}^{(s)}}$ plus rejuvenation,
  	then with probability
    \[
    \alpha = \min\left(1, \, \frac{p(\theta_{i+1}^{\ast}|\zeta^{(l)}) }{p(\theta_{i+1}^{(l-1)}|\zeta^{(l)})}\cdot \frac{p(\theta_{i+1}^{(l-1)} | \zeta_\text{ref})}{p(\theta_{i+1}^{\ast} | \zeta_\text{ref})}\right)
    \]
    set $\theta_{i+1}^{(l)} = \theta_{i+1}^{\ast}$, otherwise $\theta_{i+1}^{(l)} = \theta_{i+1}^{(l-1)}$\;
}\medskip
\emph{Parametric approximation of posterior:}\\[3mm]
estimate $H_{i+1}$ based on $\zeta^{(1:L)}$ such that 
$p(\cdot\,;H_{i+1}) \approx \sum_{l=1}^{L} \delta_{\zeta^{(l)}}$
}
\end{algorithm}

	\bigskip

	\section{Numerical experiments}

	\subsubsection*{Pharmacokinetic model}
	
	As an example model to evaluate the different algorithms, we considered a one-compartment pharmacokinetic model, with a single intravenous bolus dose $d = 100\, \mu\text{mol}$ administered at time 0 and first order elimination kinetics.
	In this model, the plasma concentration $C(t,\theta)$ predicted at time $t$ can be computed analytically:
	\begin{equation}
	\label{eq:Ct} 
		C(t,\theta) = \frac{D}{V} \exp\left(-\frac{e^{\theta} t}{V}\right),  
	\end{equation}
	with volume of distribution $V$ and log-clearance $\theta$. 
	While $V = 20$\,L is assumed to be fixed, inter-individual variability (IIV) is assumed on log-clearance,
	given by a normal distribution (corresponding to a lognormally distributed clearance)
	\begin{equation}
	\label{eq:IIV}
		\theta_i \sim \mathcal{N}(\mu_\text{CL},\omega^{2}_\text{CL}).
	\end{equation}
	Additionally, we assume random unexplained variability (RUV) on the observed data, given by an additive error on log-scale:
	\begin{equation}
	[y_{ij}|\theta_{i}] \sim \log\mathcal{N}\Big( \log\big(C(t_{ij},\theta_{i})\big), \sigma^{2}\Big),
	\label{eq:RUV}
	\end{equation}
	with error variance (on log scale) $\sigma^2$.
	That is, $\zeta = (\mu_\text{CL},\omega^{2}_\text{CL})$, with $V$, $\sigma^{2}$ being assumed to be known from the prior data.
	 
	\subsubsection*{Prior distribution}
	
	As a prior for the population parameters $\mu_\text{CL},\omega^{2}_\text{CL}$, a normal-inverse gamma distribution
	\begin{equation}
	\label{eq:prior-example}
	(\mu_{CL},\omega^{2}_{CL}) \sim \text{N-}\Gamma^{-1}(\mu_{0},\kappa_{0},\alpha_{0},\beta_{0})
	\end{equation}
	is chosen, with parameters $\mu_{0} = \log(5)$, $\kappa_{0} = 1$, $\alpha_{0} = 10$ and $\beta_{0} = 2.7$. 
	This distribution is conjugate to a normal distribution with unknown mean and variance, a necessary requirement to use Algorithm~\ref{algo:MwG} (parametric approximation + Metropolis-within-Gibbs).
	Of note, this specific choice is not required for any of the other algorithms.

	\subsubsection*{Data generation} 
	
	We considered a simulation study in order to have a scenario in which the ground truth is known and where algorithms can be compared.
	Mimicking the anticipated change of distribution between trial and real-world data,
	individual parameters $\theta_{1},...,\theta_{N}$ were assumed to originate from a distribution that is very unlikely under the prior,
	namely a normal distribution with parameters 
	\[
	\zeta_\text{true} = (\mu_\text{CL,true},\omega^{2}_\text{CL,true}) = (\log(2), 0.1),
	\]
	which corresponds to a median clearance of 2 L/h.	
	Four scenarios of different data sparsity were considered, with either $n=20$ or $n=100$ individuals, and per individual either a sparse sampling scheme with observation times $t =$ 0\,h and 1\,h,  or a rich sampling scheme with observation times $t =$ 0\,h, 1\,h, 2\,h, 5\,h, 11\,h, 23\,h and 47\,h. 
	
	\subsubsection*{Algorithmic considerations}

	For the pseudo-marginal Metropolis-Hastings algorithm, $M = 25$ Monte Carlo samples were used in 	\eqref{eq:pseudo-marginal-likelihood}, and a chain length of $L= 10^{4}$ (with 10\% burn-in). 
	Based on preliminary testing, the proposal distribution 
	\[
	q(\zeta^{*}|\zeta) = \mathcal{N}\left(\zeta,\begin{pmatrix} 0.2 & 0 \\ 0 & 0.2\end{pmatrix}\right)
	\]
	was chosen.

	For the nested and single inner nested particle filters, $S=R=1000$ outer and inner particles were used.
	For the single inner nested particle filter, the outer ensemble was summarized via the componentwise weighted median (i.e., the median of the empirical distribution $\sum_{r=1}^{R}v_{r}^{i}\delta_{\zeta_{r}^{i}}$ of the posterior approximation).
	No rejuvenation was used for either particle filter method (see also Sec.~\ref{sec:conclusion}).
	
	For the parametric approximation + Metropolis-within-Gibbs algorithm, a normal-inverse gamma distribution was assumed (a special case of the normal-inverse Wishart distribution for the multivariate case), which is conjugate to a normal distribution with unknown mean and variance, cf.~\eqref{eq:IIV} and~\eqref{eq:prior-example}.
	The representation $\zeta_\text{ref}$ was chosen to be the mean of the parametric distribution $p(\cdot\,;H_{i})$.
	As for pseudo-marginal Metropolis-Hastings, a chain length of $L= 10^{4}$ with 10\% burn-in was used.
	The parameter $H_{i+1}$ from the parametric approximation was estimated from the sample $\zeta^{(1:L)}$ resulting from the Metropolis-within-Gibbs procedure by matching theoretical and empirical mean and variance. 
	For the rejuvenation step, a normal distribution with standard deviation of 0.1\% of the sampled individual parameter $\theta_{i+1}^*$ was chosen.

	\subsubsection*{Results}

\begin{figure}[htp!]
	\centering
	\includegraphics[width=.8\textwidth]{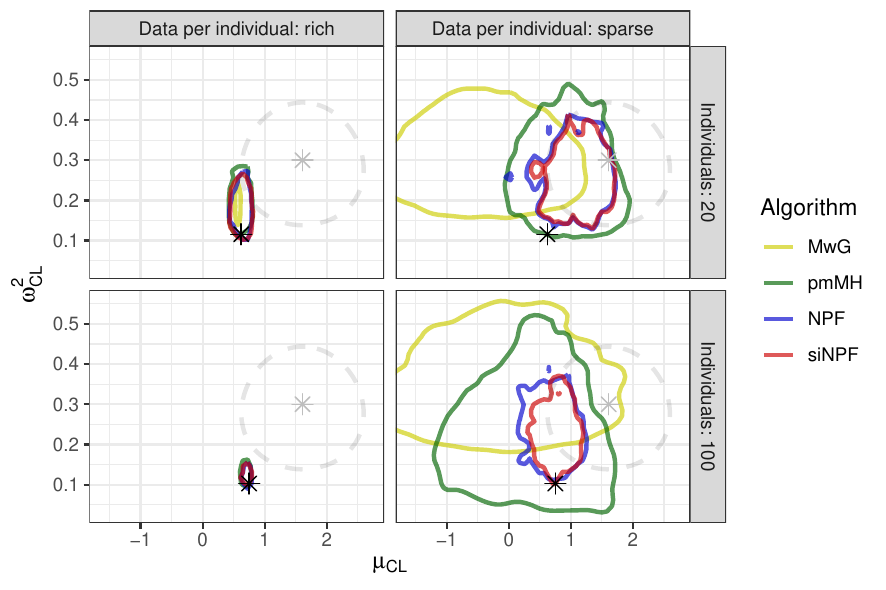}
	\caption{\label{fig:results} 
		Accuracy comparison of different algorithm for marginal posterior inference.
		The prior distribution $p(\zeta)$ is shown in grey dashed line, with its mean indicated by the grey star.
		Colored bold lines represent different approximations to the marginal posterior $p(\zeta|\mathbf{y}_{1:N})$.
		All distributions are represented as 80\% highest density regions based on kernel density estimation. 
		The data-generating parameters $\zeta_\text{true}$ are shown as a black star.\\
		\emph{Abbreviations: MwG\,=\,Metropolis-within-Gibbs \& parametric approximation; pmMH\,=\,pseudo-marginal Metropolis-Hastings with importance sampling; NPF\,=\,nested particle filter; siNPF\,=\,single inner nested particle filter.}
	}
	\end{figure}

	For each data scenario, we simulated a real-world population of individuals and 
	investigated all algorithms for hierarchical Bayesian inference.
	The results are shown in Figure~\ref{fig:results}.
	As expected, the reference method (pseudo-marginal Metropolis-Hastings with importance sampling) covers the data-generating parameters in all data sparsity scenarios, with more uncertainty under more sparse conditions.
	The two particle filters (nested and single inner nested particle filter) show a very similar behaviour to each other, consistent with the reference method in rich data scenarios, but overconfident in more sparse data scenarios.
	Finally, the parametric approximation + Metropolis-within-Gibbs algorithm is only accurate in the data-richest scenario, while it is overconfident and even biased in more sparse scenarios.

	Runtimes were compared between the four algorithms for a single batch inference on all individuals (see Table~\ref{tab:runtime}).
	Compared to the reference method (pseudo-marginal Metropolis-Hastings with importance sampling), the nested particle filter took slightly (approx.~25\%) longer,
	while the parametric approximation + Metropolis-within-Gibbs and single inner nested particle filter were 20 times and 300 times faster, respectively.
	Runtimes in different data sparsity scenarios were almost unaffected by the number of datapoints used per individual and scaled linearly with the number of individuals. 
	Therefore, the above relative runtime comparisons were independent of the data sparsity scenario.
	Comparing the nested and single inner nested particle filters, a 400 fold speedup was observed by avoiding 	the creation of 1000$\times$1000 inner particles.
	For more costly models, this accelaration factor is expected to increase roughly to the number of outer particles (here, 1000).
	Of note, in a setting where forecasting after each individual is required, the runtime cost for batch algorithms would increase considerably, by a factor roughly proportional to the number of individuals considered.

	\begin{table}[htp!]
	\centering
	\begin{tabular}{r||c|c|c|c}
	 & \multicolumn{2}{c|}{$N=20$} & \multicolumn{2}{c}{$N=100$}\\
	Algorithm & sparse & rich & sparse & rich \\	
	\hline
	pseudo-marginal Metropolis-Hastings & 1268\,s & 1190\,s & 5581\,s & --\\
	nested particle filter & 1590\,s & 1635\,s & 7904\,s & 8183\,s\\
	single inner nested particle filter & 4\,s & 4\,s & 20\,s & 21\,s\\
	parametric approximation & 82\,s & 84\,s & 422\,s & 419\,s\\
	\end{tabular}
	\caption{\label{tab:runtime}Runtime comparison of the different algorithms (MacBook Pro, 2 GHz Quad-Core Intel Core i5 processor,  16 GB 3733 MHz LPDDR4X memory).}
	\end{table}

	\section{Conclusion}\label{sec:conclusion}
	
	In this work, we evaluated the accuracy and runtime of different algorithms for hierarchical Bayesian estimation in a simplified MIPD setting.
	The pseudo-marginal Metropolis-Hastings algorithm was considered as a reference method, and it showed consistent behaviour in all scenarios.
	Three sequential algorithms were considered, a previously investigated parametric approximation + Metropolis-within-Gibbs and two variants of nested particle filters. 
	Consistent with \cite{Maier2021}, we observed that the parametric approximation + Metropolis-within-Gibbs algorithm yielded incorrect posterior distribution estimates in sparse data scenarios.
	A possible reason could be the sequential amplification of approximation errors of the posterior within the parametric class.
	The nested particle filters showed better accuracy, but still slight overconfidence in the posterior.  	
	Compared to the nested particle filter, the single inner nested particle filter resulted in very similar performance at greatly reduced runtime; overall, it had a good tradeoff between accuracy and computational cost. 
	For sampling a reference ensemble, the median was used to summarize the population distribution, which worked well in the simple model considered here. 
	In more complex (and higher-dimensional) models, a variance inflated summary might be appropriate to achieve a better parameter space coverage.
	Furthermore, the remaining misfit of particle filters might be addressed by appropriately chosen particle rejuvenation. 
	Future work will further develop these algorithms in order to carry over rejuvenated particles from the inner to the outer level.
	Finally, more complex PK models will be studied in future work, in particular the neutropenia setting investigated in \cite{Maier2021}.

	\section*{Funding note}

	The research has been funded by the Deutsche Forschungsgemeinschaft (DFG) – Project-ID 318763901 – SFB1294. 


	\bibliographystyle{plain} 

	\bibliography{Hierarchical-Bayesian-Methods} 



\end{document}